\documentclass[12pt]{article}
\usepackage[dvips]{graphicx}
\usepackage{epsf}
\def\be{\begin{equation}}
\def\ee{\end{equation}}
\def\beq{\begin{eqnarray}}
\def\eeq{\end{eqnarray}}

\begin{document}
\begin{center} 
{\large\bf Influence of microwave fields on the electron transport \\
through a quantum dot in the presence of a direct \\
tunneling between leads\\}
\vspace{1cm}
{R. Taranko\footnote[1]{corresponding author, e-mail: taranko@tytan.umcs.lublin.pl}
, T. Kwapi\'nski,  E. Taranko}

\vspace{1cm}

{Institute of Physics, M. Curie-Sk\l odowska University, \\
20-031 Lublin, Poland}
\end{center}

\noindent

\bigskip\bigskip
\begin{abstract}
We consider the time-dependent electron transport through a quantum dot 
coupled to two leads in the presence of the additional over-dot (bridge)
tunneling channel. By using the evolution operator method together with 
the wide-band limit approximation we derived the analytical formulaes for
the quantum dot charge and current flowing in the system. The influence of
the external microwave field on the time-average quantum dot charge, 
the current and the derivatives of the average current with respect to 
the gate and source-drain voltages has been investigated for a wide range 
of parameters. 
\end{abstract}

\section{Introduction}

Electronic transport in mesoscopic systems has been
at the focus of experimental and theoretical 
interest during the last decade due to recent development
in fabrication of small electronic devices and their
interesting equilibrium and non-equilibrium properties.
Especially interesting are the transport properties of 
a quantum dot (QD) under the influence of external 
time-dependent fields. The high-frequency signals may be
applied to a QD and the time-dependent fields will modify 
the tunneling current.

New effects have been observed and theoretically described,
e.g. photon-assisted tunneling through small quantum dots
with well-resolved discrete energy states [1, 2, 3], photon-electron
pumps [4, 5, 6] and others.
One can investigate the current flowing through a QD under
periodic modulation of the QD electronic structure [7] or
periodic (non-periodic) modulation of the tunneling barriers
[6] and electron energy levels in both (left and right)
electron reservoirs [8] (see also [9, 10]).

The progress of nanomaterials science has enabled 
the experimental study of the phase coherence
of the charge carriers in many mesoscopic systems.
The asymmetric Fano line shapes [11] are
observed whenever resonant and nonresonant scattering
paths interfere. In some nanostructures, e.g. in
single-electron transistors, the Fano resonances in the 
conductance were observed [12], which imply
that there are two paths for transfer of electrons
between a source and a drain. Especially the recent
experimental and theoretical study with a low-temperature 
scanning tunneling microscope (STM) of the single magnetic
atom deposited on a metallic surface showed the asymmetric
Fano resonances in the tunneling spectra
[12--15]. The STM measurements indicate that in tunneling
of electrons between STM tip and a surface with a single
impurity atom two different paths are present. The
electrons can tunnel between the tip and the adsorbate 
state and directly between the tip and the metal surface.
The electronic transport through a QD coupled to the electron
reservoirs within a model with two electron tunneling channels
was considered in Ref. [16] and it was shown, that
transport of electrons through both channels leads to
an asymmetric shape of the zero bias voltage conductance
curves, which is typical behaviour for a Fano resonance
resulting from constructive and destructive interference processes
for electrons transmitted through both channels.

In all papers mentioned above and relating to the electron 
transport through a QD with the additional (bridge) transmission channel
the external fields were not applied and the
considered systems were driven out of equilibrium only
by means of a dc
voltage bias. In this paper we address the issue of a QD
with a bridge channel between a source and a drain driven out
of equilibrium by means of a dc voltage bias and additional
time-dependent external fields. In this manner, our paper can
be seen as generalization of Ref. [9] to the case of a QD
with the additional bridge channel in the presence of external
microwave fields which are applied to the dot and two leads,
respectively. In literature, different theoretical approaches
have been developed to treat the time-dependent, nonequilibrium
electron transport 
processes in the mesoscopic systems. It seems, that the
most popular is the nonequilibrium Green's function method.
However, these Green's functions depend on the time arguments
and for non trivial quantum models it is a rather difficult task
to calculate them. In our treatment of the time-dependent 
tunneling through a mesoscopic system we use the evolution 
operator technique (e.g. [17, 18]).
The final expressions for the QD charge and the
current flowing in the system can be described
in terms of the corresponding matrix elements of
the evolution operator. In our earlier work [19]
we have considered the similar problem solving
numerically the corresponding sets of the differential
equations satisfied by the matrix elements of the
evolution operator. Due to the complexity and a large number
of these equations we considered only a very
limited number of intereting cases although we were
able to take into consideration the electronic structure of 
the lead energy bands and the specific time-dependence
of the QD-lead barriers. Here we give the analytical
expressions for the QD charge and current assuming so called
the wide-band limit approximation and the time-independent
strength of the QD-lead barriers. As a result, due to these analytical 
forms, we are able to analyze the required characteristics 
of the considered system for the very broad class of 
parameters. Additionally, due to the final forms given for
some matrix elements of the evolution operator it is possible
to build up the expressions for the QD charge or current
in the form of the perturbation series.

In the next Section we present the model and formalism
and give the resulting expressions for equations for
the corresponding matrix elements of the evolution
operator. In Section 3 we obtain the approximate solutions
for all required matrix elements and give the final
forms for the QD charge and the current flowing
in the system. The results are presented in Section 4 which includes
also the summary and a brief discussion.

\section{Model and calculation method}

We model the QD coupled to the left and right electron reservoirs with
the additional bridge tunneling channel between them by the usually used
Hamiltonian $H = H_1 + V$, where 
\beq
&& H_1 = \sum_{\vec k_\alpha} \varepsilon_{\vec k_\alpha}(t) a^+_{\vec
k_\alpha} a_{\vec k_\alpha} + \varepsilon_d(t) a^+_d a_d\,,\\
&& V = \sum_{\vec k_\alpha} V_{\vec k_\alpha d}(t) a^+_{\vec
k_\alpha}a_d + {\rm h.c.} + \sum_{\vec k_L,\vec k_R} V_{\vec k_L\vec
k_R}(t) a^+_{\vec k_L} a_{\vec k_R} + {\rm h.c.}
\eeq
The operators $a_{\vec k_\alpha}(a^+_{\vec k_\alpha})$,
$a_d(a^+_d)$   are the annihilation (creation) operators of the
electron in the lead $\alpha$ ($\alpha = L, R$) and in the QD, respectively. 
The couplings between QD and lead states and between both lead states are denoted 
by $V_{\vec k_\alpha d}$ and $V_{\vec k_L
\vec k_R}$, respectively.
For simplicity,
the dot is characterized only by a single level $\varepsilon_d$ and we have
neglected the intradot electron-electron Coulomb interaction. We assume
the case in which there exist microwave fields applied to the leads and
QD. In the adiabatic approximation our time-dependent driven system is
described by $\varepsilon_{\vec k_\alpha}(t) = \varepsilon_{\vec
k_\alpha} + \Delta_\alpha\cos\omega t$, $\varepsilon_d(t) =
\varepsilon_d + \Delta_d \cos\omega t$, i.e. the energy levels of the
leads and QD are driven by the ac field with the frequency $\omega$ and
the amplitudes $\Delta_\alpha$ and $\Delta_d$, respectively. 

We describe the dynamical evolution of the charge localised on the QD
and the current flowing in the system in terms of the time evolution
operator $U(t,t_0)$ (in the interaction representation) which satisfies the
equation 
\be
 i{\partial\over\partial t} U(t,t_0) = \tilde V(t)\,U(t,t_0)\,,
\ee
where
\beq
 && \tilde V(t) = U_0(t,t_0) \, V(t) \, U^+_0(t,t_0) \,,\\
 && U_0(t) = T\exp\left(i\int\limits^t_{t_0} dt' H_1(t')\right)\,. 
\eeq
Here we assume that the interaction between QD and leads and between
both leads is switched on in the distant past $t_0$, i.e. $V_{\vec k_\alpha
d}(t)$ and $V_{\vec k_L\vec k_R}(t)$ equal
to zero for $t \leq t_0$ and takes constant values for $t > t_0$.

The QD charge and currents flowing in the system can be obtained from
the knowledge of the appropriate matrix elements of the evolution
operator $U(t,t_0)$.
The QD charge is given as follows (cf. [18]):
\be
n_d(t) = n_d(t_0)|U_{dd}(t,t_0)|^2 + \sum_{\vec k_\alpha} n_{\vec k_\alpha}
(t_0)|U_{d\vec k_\alpha} (t,t_0)|^2\,,
\ee
where $U_{dd}(t,t_0) \equiv \langle d|U(t,t_0)|d\rangle$  and 
$U_{d\vec k_\alpha}(t,t_0) \equiv \langle d|U(t,t_0)|\vec
k_\alpha\rangle$  denote the matrix elements of $U(t,t_0)$ calculated within
the basis functions containing the electron single-particle functions
of the leads and QD, $|\vec k_L\rangle$, $|\vec k_R\rangle$ and
$|d\rangle$, respectively. $n_d(t_0)$ and $n_{\vec k_\alpha}(t_0)$ 
represent the initial filling of the corresponding single-particle
states. 

The tunneling current flowing, e.g. from the left lead into the QD and the
right lead, $j_L(t)$ , can be obtained from the time derivative of the total
number of electrons in the left lead, $j_L(t) = -edn_L(t)/dt$ (cf.
[9]), where 
\beq
n_L(t) &=& \sum_{\vec k_L} n_{\vec k_L}(t) = \sum_{\vec
k_L}(n_d(t_0)|U_{\vec k_L d}(t,t_0)|^2  \nonumber\\
&+& \sum_{\vec q_L} n_{\vec q_L}(t_0)|U_{\vec k_L,\vec q_L}(t,t_0)|^2 + \sum_{\vec
k_R} n_{\vec k_R}(t_0)|U_{\vec k_L,\vec k_R}(t,t_0)|^2)\,.
\eeq
Let us begin with the calculations of the QD charge $n_d(t)$. Then we have to
calculate the matrix elements $U_{dd}(t,t_0)$ and $U_{d\vec
k_\alpha}(t,t_0)$.  Using the identity operator
$I = |d\rangle\langle d| + \sum_{\vec k_\alpha}|\vec k_\alpha
\rangle\langle \vec k_\alpha|$ the following set of coupled equations
can be obtained from Eq. (3): 
\beq
&& {\partial\over\partial t} U_{dd}(t,t_0) = -i\sum_{\vec k_{\alpha d}}
\tilde V^*_{\vec k_\alpha d}(t) U_{\vec k_\alpha d}(t,t_0)\,,\\
&& {\partial\over\partial t} U_{\vec k_L d}(t,t_0) = -i\tilde V_{\vec
k_L d}(t) \, U_{dd}(t,t_0) - i\sum_{\vec k_R} \tilde V_{\vec k_L\vec
k_R}(t) U_{\vec k_R d}(t,t_0)\,,\\
&& {\partial\over\partial t} U_{\vec k_R d}(t,t_0) = -i\tilde V_{\vec
k_R d}(t) \, U_{dd}(t,t_0) - i\sum_{\vec k_L} \tilde V_{\vec k_R\vec
k_L}(t) U_{\vec k_L d}(t,t_0)\,,
\eeq
where 
\beq
\tilde V_{\vec k_\alpha d}(t) &\equiv& \langle \vec k_\alpha|\tilde
V(t)|d\rangle = V_{\vec k_\alpha d}\, \exp(i(\varepsilon_\alpha -
\varepsilon_d)(t-t_0) \nonumber \\
&& - i(\Delta_d - \Delta_\alpha)(\sin\omega t -
\sin\omega t_0)/\omega) \,,\\
 \tilde V_{\vec k_L\vec k_R}(t) &\equiv& \langle \vec k_L|\tilde
V(t)|\vec k_R\rangle = V_{\vec k_L\vec k_R} \exp(i(\varepsilon_{\vec k_L} -
\varepsilon_{\vec k_R})(t-t_0) - \nonumber \\
&& - i(\Delta_L - \Delta_R)(\sin\omega t - \sin\omega t_0)/\omega) \,.
\eeq
It is easy to show that the equation for $U_{dd}(t,t_0)$ can be written as
follows:
\be
 {\partial U_{dd}(t,t_0)\over\partial t} =
-\int\limits^t_{t_0}dt'({\cal K}(t,t')U_{dd}(t',t_0) + \sum_{\vec
k_\alpha} {\cal L}_{\vec k_\alpha}(t,t')\, U_{\vec k_\alpha
d}(t',t_0))\,, 
\ee
where
\beq
 {\cal K}(t,t') = \sum_{\vec k_\alpha} \,\tilde V^*_{\vec k_{\alpha d}}(t) 
\tilde V_{\vec k_{\alpha d}}(t')\,,\\
{\cal L}_{\vec k_L}(t,t') = \sum_{\vec k_R} \tilde V^*_{\vec k_R}(t) \,
\tilde V_{\vec k_R\vec k_L}(t')\,,
\eeq
and the similar equation can be written for ${\cal L}_{\vec k_R}(t,t')$.

The formal solution of Eq. (9) written in the form
\be
U_{\vec k_L d}(t,t_0) = -i\int\limits^t_{t_0} dt' \tilde V_{\vec k_L d}(t')
U_{dd}(t',t_0) - i\sum_{\vec k_R} \int\limits^t_{t_0} dt' \tilde V_{\vec
k_L\vec k_R}(t') U_{\vec k_R d}(t')\,,
\ee
(and the similar equation for $U_{\vec k_R d}(t,t_0)$) can be iterated giving


\beq
\lefteqn{
U_{L d}(t,t_0) = -i\int\limits^t_{t_0} dt_1 \, \tilde V_L(t_1)\,
U_{dd}(t_1,t_0) } \nonumber  \\
&+& \sum^\infty_{j=2} (-i)^j \sum_{R_1,L_2,R_3,L_4,\ldots,\alpha_{j-1}} 
\int\limits^t_{t_0} dt_1 \ldots \int\limits^{t_{j-1}}_{t_0} dt_j \,
\tilde V_{LR_1}(t_1) \, \tilde V_{R_1 L_2}(t_2) \ldots \tilde 
V_{\alpha_{j-1}} (t_j)\cdot U_{dd}(t_j,t_0)\,.\nonumber \\
&&
\eeq
Here we have introduced the abbreviated form for the vector $\vec k_\alpha$ and replaced it
by $\alpha$ and $\tilde V_{\vec k_\alpha d}(t) \equiv \tilde V_\alpha(t)$. The equation for $U_{R d}(t,t_0)$ can be obtained from Eq. (17) by
interchanging $L\leftrightarrow R$. 
Inserting these expressions for $U_{L d}(t,t_0)$ and $U_{Rd}(t,t_0)$
into Eq. (13) one obtains
the closed integro-differential equation for $U_{dd}(t,t_0)$
\beq
\lefteqn{
{\partial\over\partial t} U_{dd} (t,t_0) = -\int\limits^t_{t_0} dt_1 \,
 {\cal K}(t,t_1) U_{dd}(t_1,t_0) } \nonumber\\
&-&  \sum^\infty_{j=2} (-i)^{j-1} \sum_{L_1,R_2,L_3,\ldots,\alpha_j} 
\int\limits^t_{t_0} dt_1 \ldots \int\limits^{t_{j-1}}_{t_0} dt_j \,
\tilde V^*_{L_1}(t) \, \tilde V_{L_1 R_2}(t_1) \ldots \tilde 
V_{\alpha_j} (t_j)\, U_{dd}(t_j,t_0) \nonumber \\
&& + {\rm (the~second~term~with~the~change}~L\leftrightarrow R).
\eeq
Equation (18) together with the expressions for $\tilde V_L(t)$ and
$\tilde V_{LR}(t)$, Eqs. (11, 12), and for ${\cal K}(t,t_1)$ written as follows 
\be
{\cal K}(t,t_1) = \sum_{\alpha = L,R} |V_\alpha|^2 \, {\cal D}_\alpha
(t-t_1) \,\exp(i\varepsilon_d(t-t_1) + i(\Delta_d -
\Delta_\alpha)(\sin\omega t - \sin\omega t_1)/\omega)
\ee
where ${\cal D}_{\alpha}(t)$ a Fourier transform of the $\alpha$-th lead density of states,
gives exact, closed equation for $U_{dd}(t,t_0)$. Here we have assumed that 
$V_{\vec k_\alpha}$ does not depend on the wave vector $\vec k_\alpha$
and then the similar assumption will be made for $V_{\vec k_L\vec 
k_R}$.

Under the wide-band limit (WBL) approximation (e.g. [9]) this equation can be
analytically solved and such solutions will be considered later.
Formally, solving Eq. (18) and inserting its solution to Eq. (17),
the solutions for $U_{L d}(t,t_0)$ and $U_{Rd}(t,t_0)$ can be  obtained. These functions
are needed in the calculations of the first term of $j_L(t)$ (see Eq. 7). 

In order to calculate $n_d(t)$ we still need $U_{d\alpha}(t,t_0)$. Writing down the set of
closed equations for $U_{dR}(t,t_0)$, $U_{R_1 R_2}(t,t_0)$ and $U_{LR}(t,t_0)$ (obtained on the basis of Eq.
(3)) and performing similar calculations to those described above, one
obtains for $U_{dR}(t,t_0)$ (and similar equation for $U_{dL}(t,t_0)$
by interchanging $L\leftrightarrow R$):
\beq
\lefteqn{
{\partial\over\partial t} U_{dR}(t,t_0) =}\nonumber \\
&-& i\tilde V^*_R(t) + (-i)^2 \int\limits^t_{t_0}dt_1 \sum_L \tilde V^*_L(t) \tilde V_{LR}(t_1)  \nonumber \\
&+& \sum^\infty_{j=2,4,6,\ldots} (-i)^{j+1}
\sum_{R_1,L_2,R_3,\ldots,L_j} \int\limits^t_{t_0}dt_1 \ldots \int\limits^{t_{j-1}}_{t_0} dt_j
 \tilde V^*_{R_1}(t) \tilde V_{R_1L_2}(t_1) \ldots \tilde V_{L_jR}(t_j) \nonumber \\
&+& \sum^\infty_{j=3,5,7,\ldots} (-i)^{j+1} \sum_{L_1,R_2,L_3,\ldots,L_j}
\int\limits^t_{t_0}dt_1 \ldots \int\limits^{t_{j-1}}_{t_0} dt_j \, \tilde
V^*_{L_1}(t) \tilde V_{L_1R_2}(t_1) \ldots \tilde V_{L_jR}(t_j) \nonumber\\ 
&+& (-i)^2 \int\limits^t_{t_0} dt_1 \sum_L \tilde V^*_L(t) V_L(t_1)
 U_{dR}(t_1,t_0) \nonumber \\
&+& \sum^\infty_{j=2} (-i)^{j+1} \sum_{L_1,R_2,L_3,\ldots,\alpha_j} \int\limits^t_{t_0}dt_1 \ldots \int\limits^{t_{j-1}}_{t_0} dt_j \,
\tilde V^*_{L_1}(t) \tilde V_{L_1R_2}(t_1) \ldots \tilde V_{\alpha_j}(t_j) U_{dR}(t_j,t_0) \,,
\eeq
and for $U_{LR}(t,t_0)$ (needed for calculation
of the last term of $n_L(t)$, Eq. (7)):
\beq
\lefteqn{                                 
U_{LR}(t,t_0) = -i\int\limits^t_{t_0} dt_1 \, \tilde V_{LR}(t_1)} \nonumber\\
&+& \sum^\infty_{j=3,5,7\ldots} (-i)^j \sum_{R_1,L_2,R_3,\ldots,L_{j-1}} \int\limits^t_{t_0}dt_1\int\limits^{t_1}_{t_0} dt_2 
\ldots \int\limits^{t_{j-1}}_{t_0} dt_j
\tilde V_{LR}(t_1)\tilde V_{R_1L_2}(t_2) \ldots \tilde V_{L_{j-1}R}(t_j) \nonumber\\
&-& i\int\limits^t_{t_0} dt_1 \tilde V_L(t_1) U_{dR}(t_1,t_0) \\
&+& \sum^\infty_{j=2} (-i)^j \sum_{R_1,L_2,\ldots,\alpha_{j-1}} \int\limits^t_{t_0}dt_1 \int\limits^{t_1}_{t_0}dt_2 \ldots 
\int\limits^{t_{j-1}}_{t_0} dt_j \tilde V_{LR_1}(t_1) \tilde V_{R_1L_2}(t_2) \ldots \tilde V_{\alpha_j}(t_j) U_{dR}
(t_j,t_0) \,. \nonumber 
\eeq
The analytical solutions of these equations under the WBL approximation 
will be discussed in the next section. 

For calculation of $n_{\vec k_L}(t)$ one needs the functions $U_{Ld}(t,t_0)$,
$U_{LR}(t,t_0)$ and $U_{L_1L_2}(t,t_0)$. The first two functions are given in 
Eqs. (17) and (21) and
$U_{L_1L_2}(t,t_0)$ should be calculated from the set of coupled equations for
$U_{dL}(t,t_0)$, $U_{L_1L_2}(t,t_0)$  and $U_{RL}(t,t_0)$ obtained from Eq. (3).
The result is as follows:
\beq
\lefteqn{\hspace{-1cm}U_{L_1L}(t,t_0) = \delta_{L_1L} + \sum^\infty_{j=2,4,\ldots} (-i)^j \sum_{R_1,L_2,R_3,\ldots,R_{j-1}} \int\limits^t_{t_0} dt_1\int\limits^{t_1}_{t_0}dt_2 \ldots \int\limits^{t_{j-1}}_{t_0} dt_j \tilde V_{L_1R_1}(t_1) \tilde V_{R_1L_2}(t_2) \ldots \tilde V_{R_{j-1}L}(t_j) } \nonumber \\
&& - i\int\limits^t_{t_0} dt_1 \tilde V^*_{L_1}(t_1) U_{dL}(t_1,t_0) \\
&+& \sum^\infty_{j=2} (-i)^j \sum_{R_1,L_2,R_3,\ldots,\alpha_{j-1}}
\int\limits^t_{t_0} dt_1 \int\limits^{t_1}_{t_0} dt_2 \ldots
\int\limits^{t_{j-1}}_{t_0} dt_j \tilde V_{L_1R_1}(t_1) \tilde
V_{R_1L_2}(t_2) \ldots \tilde V_{\alpha_{j-1}L}(t_j)U_{dL}(t_j,t_0)\,, \nonumber
\eeq
where $U_{dL}(t,t_0)$ is given by solving Eq. (20) with the replacement $L\leftrightarrow R$.

\section{Analytical solutions in the WBL approximation}

In order to calculate the QD charge $n_d(t)$ or current
$j_L(t)$ one has to solve, in the first step, the
integro-differential equations satisfied by $U_{dd}(t,t_0)$, Eq. (18), and $U_{dR/L}(t,t_0)$,
Eq. (20). The other needed functions $U_{Ld}(t,t_0)$, $U_{L_1L_2}(t,t_0)$ and $U_{LR}(t,t_0)$ can be obtained from
 Eqs. (17, 21, 22) inserting into them
$U_{dd}(t,t_0)$, $U_{dL/R}(t,t_0)$ and performing multiple time and $\vec k$-vector integrations.
Unfortunately, it is a very difficult task
to solve the integro-differential equations and to perform these integrals in a general case when the leads are
characterized by some density of state curves. Here,
we use the WBL approximation, under which all multiple
time and $\vec k$-vectors integrals can be performed without difficulties.
The WBL approximation has been widely used in calculating of many
properties of mesoscopic systems (e.g. [2, 3, 5, 9, 10]). It is justified
under the conditions that the QD energy level linewidth
is much smaller than the bandwidth of the leads and
that the density of states and hopping matrix elements vary slowly
with energy. Furthermore, as we are not going to consider the case
of the QD energy level lying close to the edges of the leads energy band
or lying close to some singular structure present 
in the leads density of states, then application of the WBL approximation
should be fully justified. The conditions under which we perform
our calculations are satisfied in most experimental constructions of 
mezoscopic systems. As a check, we have performed the direct
but time consuming numerical integration of Eqs. (8--10)
(and similar equations for other functions) for the rectangular
leads density of states and did not find any differences in the results for 
the time-averaged QD charge or the currents flowing in the considered system.
Therefore, in this section we consider the electron transport
through the QD with the additional bridge tunnel over the dot
within the WBL approximation. In this approach the solutions
of the integro-differential Eqs. (19) and (20) in the analytical form
can be obtained and the infinite sums of all terms 
in Eqs. (17) and (21) can be calculated.

Let us consider the equation for the matrix element $U_{dd}(t,t_0)$,
Eq. (18). The function ${\cal K}(t,t')$, Eq. (19), is approximated 
in WBL as follows
\beq
\lefteqn{{\cal K}(t,t_1) = \sum_\alpha |V_\alpha|^2
\int^\infty_{-\infty} d\varepsilon {\cal D}_{\alpha}(\varepsilon) \,
\exp(-i\varepsilon(t - t_1)) \,\exp(-i\varepsilon_d(t-t_1)}
\nonumber \\
&& \hspace{5cm}  + i(\Delta_d - \Delta_\alpha)(\sin\omega t - \sin\omega
t_1)/\omega) \Rightarrow  \nonumber \\ 
&\Rightarrow & \sum_\alpha {|V_\alpha|^2\over D_\alpha} 2\pi\delta(t-t_1)\,,
\eeq
where the leads density of states ${\cal D}_\alpha(\varepsilon)$ was 
replaced by the  rectangular  density of states 
with the ''effective'' bandwidth $D_\alpha$

Using similar approximation in calculations of the
multiple integrals present in Eq. (18) one obtains
\be
{\partial\over\partial t}U_{dd}(t,t_0) = \left(-{\Gamma\over 2} -
{2V^2\pi\over D} \sum^\infty_{j=1} (-i\pi V_{RL}/D)^j\right)
U_{dd}(t,t_0)\,, 
\ee
where $\Gamma_\alpha = 2\pi V^2/D$, $\Gamma = \Gamma_L +\Gamma_R$, $D_L = D_R = D$, $V \equiv V_\alpha$.

Assuming $\pi V_{RL}/D < 1$ the series can be summed up and finally 
the equation for $U_{dd}(t,t_0)$ reads
\be
{\partial\over\partial t} U_{dd}(t,t_0) = -C_1 U_{dd}(t,t_0)\,,
\ee
where $C_1 = (2\pi V^2/D)/(1 + i\pi V_{RL}/D)$.

Similarly, Eq. (20) becomes
\be
{\partial\over\partial t} U_{d\alpha}(t,t_0) = 
-i\tilde V_{dd}(t)/(1 + iV_{RL} \pi/D) - C_1 U_{d\alpha}(t,t_0)\,.
\ee
The solutions of Eqs. (25) and (26) are as follows:
\beq
&& U_{dd}(t,t_0) = \exp(-C_1(t-t_0))\,,\\
&& U_{d\alpha}(t,t_0) = -i/(1 + iV_{RL}\pi/D) \int\limits^t_{t_0} dt_1
\tilde V_{d\alpha}(t_1)\,\exp(-C_1(t-t_1))
\eeq
and the QD charge $n_d(t)$ can be easily obtained, Eq. (6).
It can be verified, that the first term of Eq. (6) tends to
zero as $t - t_0 \rightarrow\infty$ and finally for the QD charge we have
\be
n_d(t) = {1 \over (1 + (V_{RL}\pi/D)^2)} \sum_\alpha {\Gamma_\alpha\over 2\pi}
\int d\varepsilon\, f_\alpha(\varepsilon)|A_\alpha(\varepsilon,t)|^2\,,
\ee
where
\be
A_\alpha(\varepsilon,t) = -\int\limits^t_{t_0}dt_1\, 
\exp(i(\varepsilon_d - \varepsilon)(t-t_1)  -
i(\Delta_d - \Delta_\alpha) (\sin\omega t - \sin\omega t_1)/\omega -
C_1 (t - t_1)) 
\ee

In the limit of vanishing bridge over the QD, Eq. (29) reproduces
the result of Ref. [9].

The current $j_L(t)$ flowing from the left lead into 
the QD and the right lead is calculated from the
evolution of the total number of electrons in the left
lead   (see Eq. 7) and one can read:
\beq
j_L(t) &=& 2 {\rm Re}\left\{n_d(t_0) \sum_{\vec k_L} U^*_{\vec k_Ld}(t,t_0) {d\over dt}
U_{\vec k_Ld}(t,t_0) \right. \nonumber \\ 
&+& \sum_{\vec k_L \vec q_L} n_{\vec q_L}(t_0) U^*_{\vec k_L,\vec q_L}
(t,t_0) {d\over dt} U_{\vec k_L,\vec q_L}(t,t_0) \nonumber \\ 
&+&\left. \sum_{\vec k_L,\vec k_R} n_{\vec k_R}(t_0) U^*_{\vec k_L, 
\vec k_R}(t,t_0) {d\over dt} U_{\vec k_L,\vec k_R}(t,t_0) \right\}\,.
\eeq
The functions $U_{\vec k_Ld}$, $U_{\vec k_L,\vec q_L}$ and $U_{\vec k_L, \vec 
k_R}(t,t_0)$ are calculated according to 
Eqs. (17), (22) and (21), respectively, and after summing up
of the corresponding multiple integrals one obtains:
\beq
U_{\vec k_Ld} (t,t_0) &=& -{i\over 1 + ix} \int\limits^t_{t_0}dt_1 \tilde
V_{\vec k_Ld}(t_1) U_{dd}(t_1,t_0)\,,\\ 
U_{\vec k_L,\vec q_L}(t,t_0) &=& \delta_{\vec k_L,\vec q_L} - {V_{RL}
\cdot x\over 
1 + x^2} \int\limits^t_{t_0} dt_1\, \exp \{i(\varepsilon_{\vec k_L} -
\varepsilon_{\vec q_L}) (t_1-t_0)\} \nonumber\\ 
&-& {i\over 1 + ix} \int\limits^t_{t_0} dt_1 \tilde V_{\vec k_Ld}(t_1)
U_{d\vec q_L}(t_1,t_0)\,,\\ 
U_{\vec k_L,\vec k_R}(t,t_0) &=& -{i\over 1 + ix}\int\limits^t_{t_0}
dt_1\,\tilde V_{\vec k_Ld}(t_1) U_{d\vec k_R}(t_1,t_0) \nonumber\\ 
&-& {i\over 1 + x^2} \int\limits^t_{t_0} dt_1 \tilde V_{\vec k_L\vec
k_R}(t_1) \,, 
\eeq
where $U_{dd}(t,t_0)$, $U_{d\vec k_L}(t,t_0)$ and $U_{d\vec k_R}(t,t_0)$ 
are given in Eqs. (27, 28) and $x = \pi V_{RL}/D$.
One can verify, that the first term in Eq. (31)
tends to zero for $t - t_0 \rightarrow\infty$ as we have for this term
\be
n_d(t_0) |1/(1 + ix)|^2 \,\Gamma/2\, \exp(-\Gamma(t - t_0) \, Re C_2)\,,
\ee
where ${\it Re} C_2 = {\it Re}(1/(1 + ix)) > 0$.

The second and third terms of the expression for $j_L(t)$, Eq. (30),
together with Eqs. (27, 28, 31-33) give finally the time averaged current:
\beq
\lefteqn{
\langle j_L(t)\rangle = {1\over\pi} {2x^2\over (1 + x^2)^2} \, 
\int (f_R(\varepsilon) - f_L(\varepsilon)) d\varepsilon + {\Gamma/2\over 
 1 + x^2} \langle n_d (t)\rangle } \nonumber \\
&& + {\rm Im} \left\{{1 - x^2\over (1 + x^2)(1 + ix)^2} {\Gamma_L\over \pi}
\int d\varepsilon f_L(\varepsilon) \langle A_L(\varepsilon,t)\rangle\right\} 
\nonumber \\
&& + {\rm Re} \left\{{2x\over (1+x^2)(1 + ix)} {\Gamma_L\over \pi} 
\int d\varepsilon f_R(\varepsilon) \langle A_R(\varepsilon,t)\rangle\right\}\,.
\eeq
In the vanishing bridge channel case Eq. (35) coincides
with the results of Ref. [9]:
\be
\langle j^{V_{RL}=0}_L(t)\rangle = {\Gamma\over 2} \langle n_d(t)\rangle + 
{\Gamma_L\over \pi} \int d\varepsilon \,f_L(\varepsilon) {\rm Im}\langle 
A_L(\varepsilon, t)\rangle\,.
\ee
The current $\langle j_L(t)\rangle$, Eq. (36), flowing from the left lead to the
central region and to the right lead (through the bridge channel)
is the superposition of four terms. The first term
corresponds to the current between two leads and this
term is not disturbed by the QD. The form of the second and
third terms is the same as for $\langle j^{V_{RL}=0}_L(t)\rangle$, 
Eq. (37), except for the renormalization constants due to the
additional tunneling channel. Note, that some additional renormalizations
also occur
due to $V_{RL}$ which enters into the expression for $n_d(t)$, Eq. (29),
and for $A_\alpha(\varepsilon,t)$, Eq. (30). The last term of Eq. (36)
is the interference term due to the simultaneous
tunneling through two channels.

\section{Results and discussion}

Here we show the numerical results of the time-averaged QD charge $\langle n_d(t)\rangle$
and the current $\langle j_L(t)\rangle$ and its derivatives with
respect to the QD energy level position and the chemical
potential $\mu_L$ (or equivalently, with respect to the
gate and source-drain voltages) for different sets
of parameters which characterize our system.
We assume the temperature $T = 0$~K and the time-average
of time-dependent quantities $f(t)$ is defined by
\be 
 \langle f(t)\rangle = \lim_{2\tau\rightarrow\infty} {1\over 2\tau} 
\int\limits^\tau_{-\tau} dt' f(t')\,,
\ee
and because $f(t)$ is a periodic function of time, 
we average it over the period $2\pi/\omega$. We take the 
chemical potential of the right lead $\mu_R$ as the energy
measurement point, $\mu_R = 0$. As the potential drop between
the left and right leads is given by $\mu_L - \mu_R = eV_{s-d}$ and
$V_{s-d}$ is the measured voltage between a source and a drain, 
then the derivatives of the current $\langle j_L(t)\rangle$ with respect to
$\mu_L$ will correspond to the derivatives $d\langle j_L(t)\rangle/dV_{s-d}$ usually
measured in experiments. 
In experiments the gate voltage controls the position of the
QD energy level $\varepsilon_d$ (regardless how complicated the 
relation between the gate voltage and $\varepsilon_d$ is) and for
that reason to mimic measurements of the QD charge 
or current vs. the gate voltage we have calculated them
vs. the position of the QD energy level $\varepsilon_d$.

The values of the hybridization matrix elements $V_{\vec k_\alpha d}$ present
in the Hamiltonian do not enter the final expressions
for the current or QD charge obtained within the WBL approximation.
Usually the effective linewidth $\Gamma_\alpha = 2\pi\Sigma_{\vec k_\alpha}
|V_{\vec k_\alpha d}|^2 \delta(E - \varepsilon_{\vec k_\alpha})$ 
is introduced. However, in our calculations the others hybridization 
matrix elements appear, $V_{\vec k_R,\vec k_L} \equiv V_{RL}$, responsible for the additional
tunneling channel for which we should take some values in order to perform 
numerical calculations. We have taken the values comparable
with $V_{\vec k_\alpha d}$ and estimated $V_{\vec k_\alpha d}$ 
(assuming its $\vec k$-independence, $V_{\vec k_\alpha d} \equiv V_\alpha = V$)
using the relation $\Gamma_\alpha = 2\pi|V_\alpha|^2/D_\alpha$, where 
$D_\alpha$ is the $\alpha$-lead's bandwidth and $D_\alpha = 100~\Gamma_\alpha$
($\Gamma_L = \Gamma_R = \Gamma$, $D_L = D_R = D$
was assumed). 
We assumed the amplitude of the QD energy levels oscillation $\Delta_d$ to be one half
of $\Delta_L$ and $\Delta_R = 0$, if otherwise stated.
In our calculations the values $V_{RL}$ were taken from the range 
(0--10), in $\Gamma$ units.

In the first three figures, Figs. 1--3, we present the overall shape of the average 
current $\langle j_L(t)\rangle$ and the derivatves of the
average current with respect to the QD energy level $\varepsilon_d$ and
the left chemical potential $\mu_L$, respectively, against
$\varepsilon_d$ and $\mu_L$. The upper panels correspond to the $V_{RL} = 0$
case and the lower ones present the results obtained
for the non-vanishing over-dot tunneling channel, $V_{RL} \neq 0$. 
There are quite visible differences between the case of a QD with and without 
additional bridge tunneling channel (cf. upper and lower panels in Figs 1--3).
For better visualization of the peculiarities of the presented functions
and for simpler discussion let us consider the specific cuts of
the surfaces given in Figs. 1-3. 

At first, let us consider
the dependence of $\langle j_L(t)\rangle$ vs. $\varepsilon_d$ for given values of the
left lead chemical potential $\mu_L$. In Fig. 4 we present
such curves for different values of $\mu_L$ -- the subsequent curves
beginning from the lower one correspond to $\mu_L = -4$ up to
the upper curve (with the step $\Delta\mu_L = 1$) obtained for $\mu_L = 8$.
The left (right) panel corresponds
to $V_{RL} = 0$ ($V_{RL} = 10$). In the case of vanishing over-dot tunneling
channel (the left panel) the current has a simple structure --
a single peak localized in the middle between $\mu_L$ and $\mu_R$
for smaller values of $\mu_L$. The width of this peak increases with 
increasing $|\mu_L|$ ($\mu_R = 0$) and for greater values of $|\mu_L|$ the current
is almost independent of $\varepsilon_d$ localized inside the energy region
between $\mu_R$ and $\mu_L$.
For non-vanishing  over-dot tunneling (Fig. 4, the right panel)
the curves $\langle j_L(t)\rangle$ become asymmetric. 
With increasing source-drain bias, the current
possesses greater values
in comparison with the $V_{RL} = 0$ case due to the
direct tunneling between both leads. Note, however , that due to the 
interference effects the resulting 
$\langle j_L(t)\rangle$ curves are asymmetric. The interference effects
are most visible approximately for $\varepsilon_d$ lying in the region ($\mu_R , \mu_L$).

In Fig. 5 we show the average current $\langle j_L(t)\rangle$ vs.
the left lead chemical potential $\mu_L$ for several
values of $\varepsilon_d$. For vanishing $V_{RL}$ the corresponding
curves are nearly asymmetric with the asymmetry point
$\mu_L = \varepsilon_d$. With increasing $\mu_L$ at fixed $\varepsilon_d$ the current
achieves a constant value depending on the position of the QD energy level
$\varepsilon_d $ with respect to the $\mu_R = 0$. 
It means, that electrons which occupy the lead energy levels not
too distant from $\varepsilon_d$ take part in the tunneling process. 
For greater $\mu_L$ most of
the lead energy levels lying far away from the $\varepsilon_d$
are inactive in the tunneling between leads through 
the QD energy level. However, at non-vanishing $V_{RL}$
(see the right panel of Fig. 5) these lead energy levels are 
active and the current $\langle j_L(t)\rangle$ vs. $\mu_L$ 
is of much richer structure. The current is nearly linearly growing
with the increasing $\mu_L$ (for larger $\mu_L$) because the tunneling through
the QD can be neglected compared with the direct
tunneling between both leads. The clearly visible
interference effects appear only for $\mu_L$ not too
distant from $\varepsilon_d$.

In Fig. 6 we show the derivatives of the average current
vs. the QD energy level $\varepsilon_d$ obtained for some values of $\mu_L$.
These are the results of the intersection of the surfaces
given in Fig. 2 with the planes at constant values of $\mu_L$ or
the results of the differentiation of curves shown in Fig. 4.
Again, the most visible differences between the results obtained
for $V_{RL} = 0$ and $V_{RL} \neq 0$ are present for the QD energy level
$\varepsilon_d$ localized approximately between chemical
potentials of both leads (compare, for example, the curves calculated
for $\mu_L = 8$).

Fig. 7 presents the comparison of the $d\langle j_L(t)\rangle/d\mu_L$ vs.
$\mu_L$ curves calculated for vanishing $V_{RL}$ (left panels)
and for $V_{RL} = 10$ (right panels) for two different values
of the amplitudes $\Delta_L$ ($\Delta_d = \Delta_L/2$, $\Delta_R = 0$).
At the vanishing value of $V_{RL}$, the
shape of the curves is symmetrical in relation
to the values $\mu_L = \varepsilon_d$ although for greater $\Delta_L$
some shoulders appear on both sides of the
corresponding peaks in the distance $\sim \Delta_L/2$ from the curve
centres. For non-vanishing $V_{RL}$, the corresponding
curves are approximately asymmetric and for
large $\mu_L$ they tend to constant, non-zero values
corresponding to linear increasing of the current at
large $\mu_L$. It is interesting that with the increasing amplitudes $\Delta_L$
and $\Delta_d$ very clear structures appear on 
both sides of the corresponding curves. Note, that
all these curves can be obtained, for example, from
the one calculated for $\varepsilon_d = 0$ and moved along
the $\mu_L$-axis by the corresponding value (equal to $\varepsilon_d$).

Fig. 8 shows the dependence of the average current $\langle j_L(t)\rangle$
on the QD energy level $\varepsilon_d$ and the direct coupling $V_{RL}$
between both leads. The upper and bottom panels
correspond to different values of the amplitude
$\Delta_L$ ($\Delta_d = \Delta_L/2$). A very distinct transition from
the symmetric to nearly antisymmetric behaviour
of the current vs. $\varepsilon_d$ curves is observed with
the increasing value of the over-dot additional
coupling between the leads for both values of $\Delta_L$.
The larger amplitudes of the left
lead and QD levels oscillations result only in some
broadening of the characteristic features of the
average current vs. $V_{RL}$ and $\varepsilon_d$ surface and do
not introduce any additional structures on these surfaces (for the range
of parameters where $\omega \sim \Gamma$ and $\mu_L-\mu_R$ is not very small).

Fig. 9 presents the average current $\langle j_L(t)\rangle$ obtained
for the case in which  only the QD energy
level $\varepsilon_d$ oscillates with some frequency ($\omega > \Gamma$ )
and for the small source-drain
voltage $\mu_L-\mu_R = 0.2$ 
. The subsequent panels a, b, c and d
correspond to the increasing value of the amplitude $\Delta_d$ and
the different curves (broken, thin, thick and very thick) 
describe the QD without the over dot channel, $V_{RL} = 0$, and with
this channel at $V_{RL} = 4,7$ and 10, respectively.
For vanishing $V_{RL}$ (broken curves) we observe for small
amplitude $\Delta_d$ only a central resonant peak (Fig. 9a). 
With increasing $\Delta_d$, the subsequent 
peaks appear and the distance between them
and the central peak is an integer multiple of the frequency $\omega$ (sidebands).
The location of peaks is independent of the
amplitude $\Delta_d$ but their relative intensity values change and
with increasing $\Delta_d$ the height of the central peak
is reduced. For the relatively large amplitude $\Delta_d$,
the heights of the two neighbouring peaks (at $\varepsilon_d \pm \omega$)
are approximately equal to the height of the 
central peak. If we take the
additional over-dot tunneling channel into consideration,
especially for small $\Delta_d$, the asymmetric shape of the
current curve is observed and this asymmetry increases with the
increasing strength of the over-dot coupling
between both leads. With the increasing amplitude
$\Delta_d$ this asymmetry is reduced largely due to the
extra, photon-assisted tunneling peaks whose strength 
increases with the increasing $\Delta_d$.
For sufficiently large values of $\Delta_d$ and $V_{RL}$
the functional dependence of the average current
on the QD energy level $\varepsilon_d$ is nearly the same
for vanishing and non-vanishing
over-dot tunneling channels. There is only one difference --
for large $V_{RL}$ the corresponding curve is shifted
to the higher values due to the direct channel between
both leads.

The last two Figures 10, 11 are devoted to the analysis of the
average current $\langle j_L(t)\rangle$ dependence on the oscillation
period $2\pi/\omega$ of the external fields. In Fig. 10 we show
the overall dependence of $\langle j_L(t)\rangle$ on $2\pi/\omega$ and the QD
energy level $\varepsilon_d$. The upper (lower) part of the Figure
presents the results for the vanishing (non-vanishing)
over dot channel between both leads. 
The most visible differences between averaged currents calculated for $V_{RL}=0$
and $V_{RL} \neq 0$ can be observed for $\varepsilon_d \leq \mu_L$, especially
for small values of $2\pi/\omega$. More
detailed analysis of the $\langle j_L(t)\rangle$ dependence on the oscillation
period $2\pi/\omega$ is shown for some chosen values of the parameters $\varepsilon_d$ and
$V_{RL}$ in Fig. 11. The thin (thick) curves correspond
to $\varepsilon_d = 5$ ($\varepsilon_d = 1$) and the solid (broken) curves
correspond to $V_{RL} = 4$ ($V_{RL} = 0$). Additionally, we
give the resutls for two values of the amplitude $\Delta_L$ 
($\Delta_d = \Delta_L/2$, $\Delta_R = 0$), i.e. for $\Delta_L = 5$ and $\Delta_L = 10$, the left
and the right parts of Fig. 11 respectively.
We observe the characteristic average current oscillations
damped with increasing $2\pi/\omega$ for $\varepsilon_d$ lying in the
central part between the left and right chemical
potentials (see also [9]). These oscillations are present for both
$V_{RL} = 0$ and $V_{RL} \neq 0$ and more visible
for greater amplitudes $\Delta_L$ and $\Delta_d$ but 
the maxima and minima of the oscillating average
current are localized at the same value of the
oscillation period. Note, that the existence of the additional over-dot
tunneling channel results approximately in shifting the
corresponding curves to higher values without any additional serious
modifications. For the QD energy level lying
away from the middle point between the left and right
chemical potentials, the average current is still an 
oscillating function of $2\pi/\omega$ although these
oscillations are less transparent and their oscillation
period is much longer. 

To conclude, we have provided a detailed investigation of a QD
connected to two leads with an additional over-dot
tunneling channel. The harmonic external microwave fields were
considered as applied to the QD and two leads which result in 
time-dependence of the corresponding QD and leads energy levels.
The QD charge and the average current flowing in this system were
calculated within the evolution operator technique.  The
corresponding matrix elements of the evolution operator
required for the calculation of the QD charge and current were
presented in the form of the infinite series of multiple time
integrals of the functions containing the information about
the coupling between the QD and leads or in the form of the
integro-differential equation.  Applying the WBL aproximation we
were able to obtain all required evolution operator matrix
elements in closed forms and give the final analytical
expressions for the QD charge and current. We have performed the
extended numerical calculations for the QD charge, the average
current and the derivatives of the current with respect to the gate
and source-drain voltages (Fig. 4).
The most spectacular influence of the additional bridge tunneling
channel is visible in the $\langle j_L(t)\rangle$ dependence vs. 
the position of the QD energy level at the constant source-drain voltage. 
Going from the vanishing values of the over-dot tunneling 
channel strength to the non-vanishing one due to the interference 
effects, the corresponding curve transforms from the symmetric to 
nearly antisymmetric shape. Similar influence of the non-vanishing
$V_{RL}$ is visible in the dependence of 
$d\langle j_L(t)\rangle/d\mu_L$ vs. $\mu_L$ (Fig. 7). The characteristic 
behaviour of the average current vs. the QD energy level position 
at the small source-drain voltage is observed for the case when the external
oscillating field is applied only to the QD (Fig. 9). For the vanishing over-dot
tunneling channel at the small amplitude $\Delta_d$, the main resonant 
peak is only observed and with the increasing amplitude $\Delta_d$ the next
peaks localized at $\varepsilon_d$ equal to the multiplicity of $\omega$
appear corresponding to the photon-assisted tunneling. In that case the current is
a symmetric curve centered around the main resonant peak.
For the nonvanishing over-dot tunneling channel and small amplitudes $\Delta_d$,
the character of the dependence of the average current on the QD energy
level position transforms with increasing $V_{RL}$ from the symmetrical to nearly 
antisymmetrical behaviour. This tendency to the antisymmetrical behaviour 
with increasing $V_{RL}$ at small amplitude $\Delta_d$ is reduced with 
increasing $\Delta_d$. For sufficiently large amplitude $\Delta_d$
the overall behaviour of the average current vs. $\varepsilon_d$ is
very similar for different values of $V_{RL}$ and does not manifest 
the tendency for the antisymmetry with increasing $V_{RL}$.

\noindent
\vspace{2cm} 

{\large\bf Literature}

\begin{enumerate}
\item T. H. Oosterkamp, L.P. Kouwenhoven, A.E.A. Koden, N.C. van der Vaart,
C.J.P.M. Harmans, Phys. Rev. Lett. {\bf 78}, 1536 (1997) 
\item Qing-feng Sun, Tsung-han Lin, Phys. Rev. {\bf B56} 3591 (1997)
\item Qing-feng Sun, Jian Wang, Tsung-han Lin,  Phys. Rev. {\bf B58},
13007 (1998) 
\item C.A. Stafford, N.S. Wingreen, Phys. Rev. Lett. {\bf 76}, 1916 (1996)
\item Qing-feng Sun, Tsung-han Lin, J. Phys.: Condens. Matter {\bf 9}, 3043 
(1997) 
\item L.P. Kouwenhoven, A.T. Johnson, N.C. van der Vaart, A. van der Enden, C.J.P.M. Harmans, 
C.T. Foxon, Z. Phys. B, Condens. Matter {\bf 85}, 381 (1991) 
\item H.-K. Zhao, J. Wang, Eur. Phys. J. {\bf B9}, 513 (1999)
\item Y. Goldin, Y. Avishai, Phys. Rev. {\bf B61}, 16750 (2000) 
\item A.-P. Jauho, N.S. Wingreen, Y. Meir,  Phys. Rev. {\bf B50}, 
5528 (1994) 
\item Qing-feng Sun, Tsung-han Lin, J. Phys.: Condens. Matter {\bf 9}, 4875 (1998) 
\item U. Fano, Phys. Rev. {\bf 124}, 1866 (1969)
\item J. G\"ores, D. Goldhaber-Gordon, S. Heemeyer, M.A. Kastner, H. Shtrikman,
D. Mahalu, U. Meirav, Phys. Rev. {\bf B62}, 2188 (2000)
\item M. Plihal, J.W. Gadzuk, Phys. Rev. {\bf B63}, 085404 (2001)
\item V. Madhavan, W. Chen, T. Jamneala, M.F. Crommie, N.S. Wingreen, Science {\bf 280},
567 (1998)
\item O. Ujsaghy, J. Kroha, L. Szunyogh, A. Zawadowski, Phys. Rev. Lett. {\bf 85},
2557 (2000)
\item B.R. Bu\l ka, P. Stefa\'nski, Phys. Rev. lett. {\bf 86}, 5128 (2001)
\item M. Tsukada, N. Shima, in: ``Dynamical Processes and Ordering on 
Solid Surfaces'', Eds. A. Yoshimori and M. Tsukada (Springer, Berlin 1995), p. 34
\item T.B. Grimley, V.C.J. Bhasu, K.L. Sebastian, Surf. Sci. {\bf 121},305 (1983)  
\item T. Kwapi\'nski, R. Taranko, Physica E, in press
\end{enumerate}


\newpage
\begin{figure}
\epsfysize=10cm
\includegraphics[angle=0,width=0.9\textwidth]{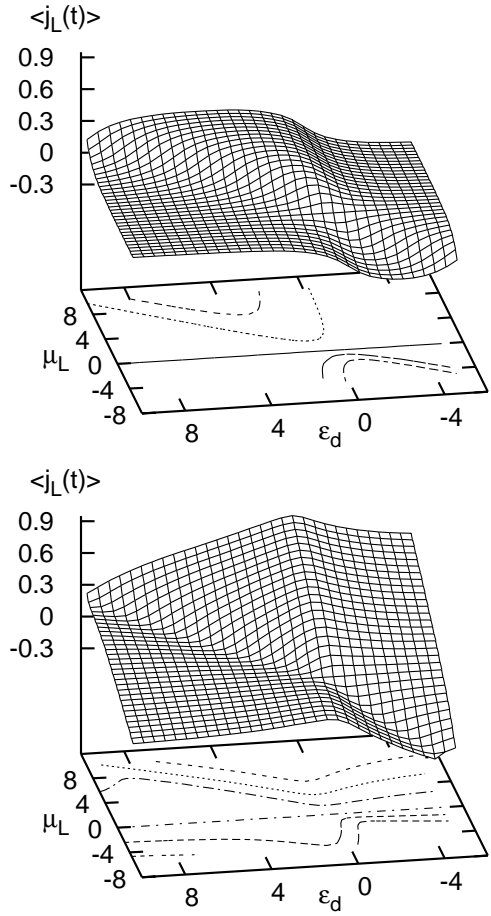}
\caption {The average current $\langle j_L(t)\rangle$ against the left lead 
chemical potential $\mu_L$ and QD energy level $\varepsilon_d$
for $V_{RL} = 0$ (the QD without the over-dot channel, the upper panel)
and for $V_{RL} = 10$ (lower panel). $\mu_R = 0$, $V = 4$, $\Delta_L = 2$, 
$\Delta_d = 1$, $\Delta_R = 0$, $\omega = 2$ and all energies are 
given in $\Gamma$ units.}
\label{1}
\end{figure}

\newpage
\begin{figure} 
\epsfysize=10cm
\includegraphics[angle=0,width=0.9\textwidth]{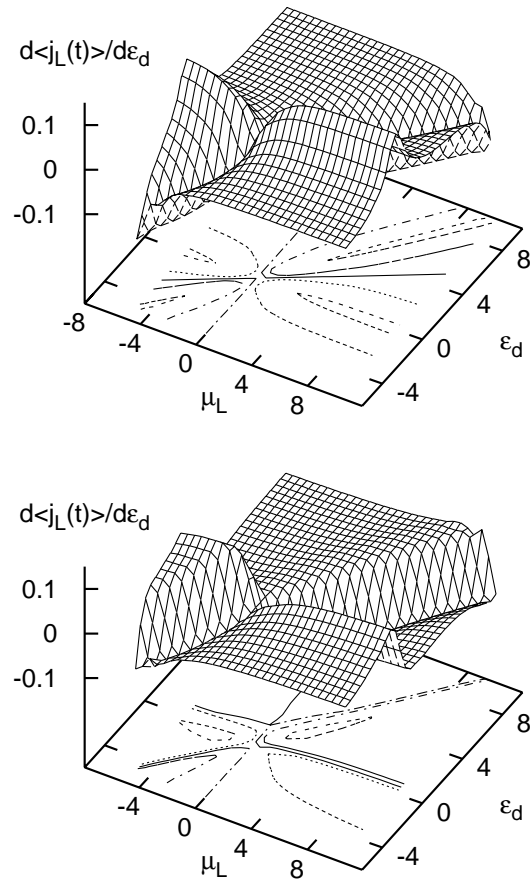}
\caption {The derivatives of the average current with respect to the QD energy level
$\varepsilon_d$, $d\langle j_L(t)\rangle /d\varepsilon_d$, against $\mu_L$ and 
$\varepsilon_d$ for $V_{RL} = 0$ (upper panel)
and for $V_{RL} = 10$ (lower panel). The other parameters as in Fig. 1.}
\label{2}
\end{figure}

\newpage
\begin{figure} 
\epsfysize=10cm
\includegraphics[angle=0,width=0.9\textwidth]{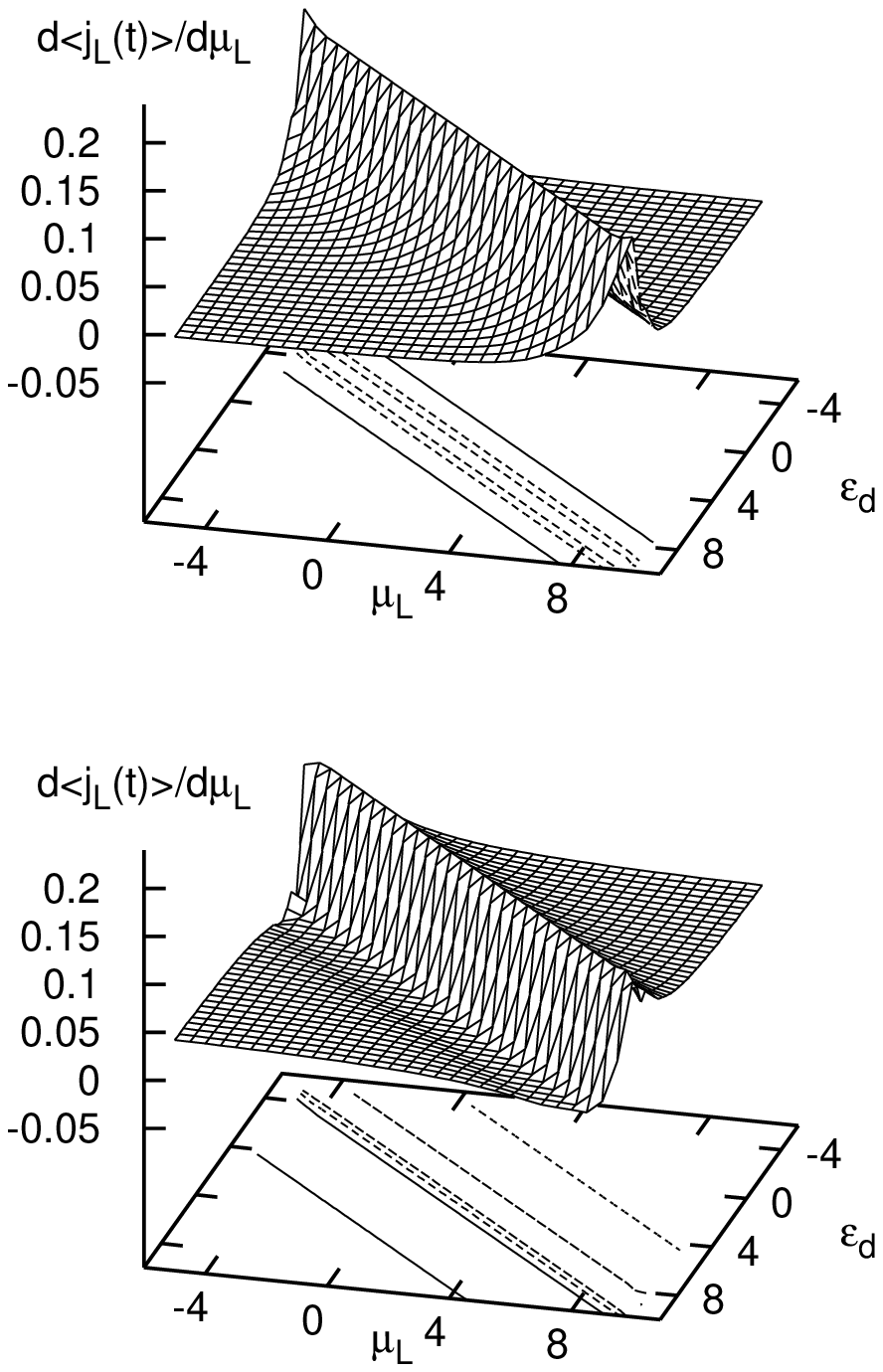}
\caption {The derivatives of the average current 
 $d\langle j_L(t)\rangle/d\mu_L$, against $\mu_L$ and 
$\varepsilon_d$ for $V_{RL} = 0$ 
(upper panel) and for $V_{RL} = 10$ (lower panel). The other parameters
as in Fig.~1.}
\label{3}
\end{figure}

\newpage
\begin{figure} 
\epsfysize=10cm
\includegraphics[angle=270,width=1.0\textwidth]{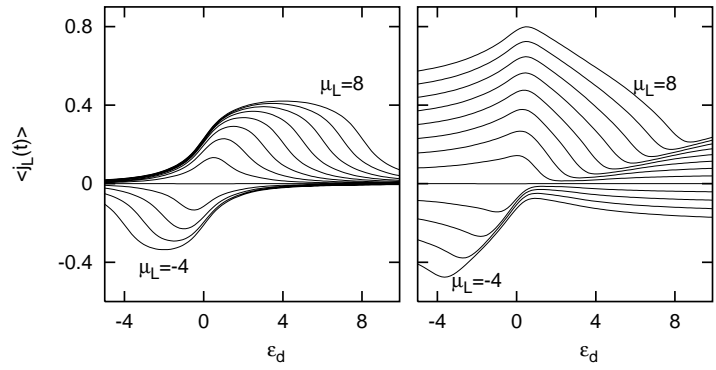}
\caption {The average current $\langle j_L(t)\rangle$ against $\varepsilon_d$ 
for given values of $\mu_L$ (beginning from $\mu_L = -4$ up to $\mu_L = 8$). The left
and right panels correspond to $V_{RL} = 0$ and $V_{RL} = 10$, respectively,
and the other parameters as in Fig. 1.}
\label{4}
\end{figure}

\newpage
\begin{figure} 
\epsfysize=10cm
\includegraphics[angle=270,width=1.0\textwidth]{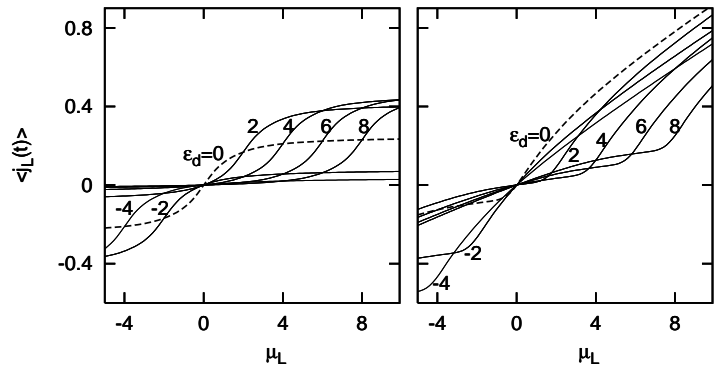}
\caption {The average current $\langle j_L(t)\rangle$ against $\mu_L$ for given
values $\varepsilon_d$ (beginning from $\varepsilon_d = -4$ up to $\varepsilon_d = 8$).
The broken curves correspond to $\varepsilon_d = 0$.
The left (right) panel corresponds to $V_{RL} = 0$ ($V_{RL} = 10)$.
The other parameters as in Fig. 1.}
\label{5}
\end{figure}

\newpage
\begin{figure} 
\epsfysize=10cm
\includegraphics[angle=270,width=1.0\textwidth]{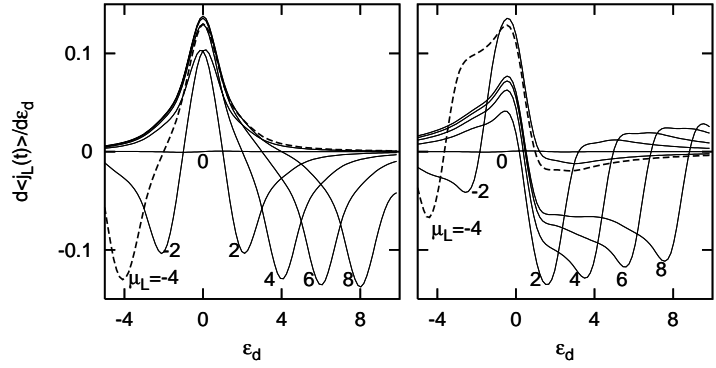}
\caption {The derivatives of the average current $d\langle j_L(t)\rangle/d\varepsilon_d$ 
with respect to the QD energy
level $\varepsilon_d$  for given values of $\mu_L$
(beginning from $\mu_L = -4$ up to $\mu_L = 8$). The broken curves correspond
to $\mu_L = -4$. The left  (right) panel
corresponds to $V_{RL} = 0$  ($V_{RL} = 10$) and the other
parameters as in Fig. 1.}
\label{6}
\end{figure}

\newpage
\begin{figure} 
\epsfysize=10cm
\includegraphics[angle=270,width=0.9\textwidth]{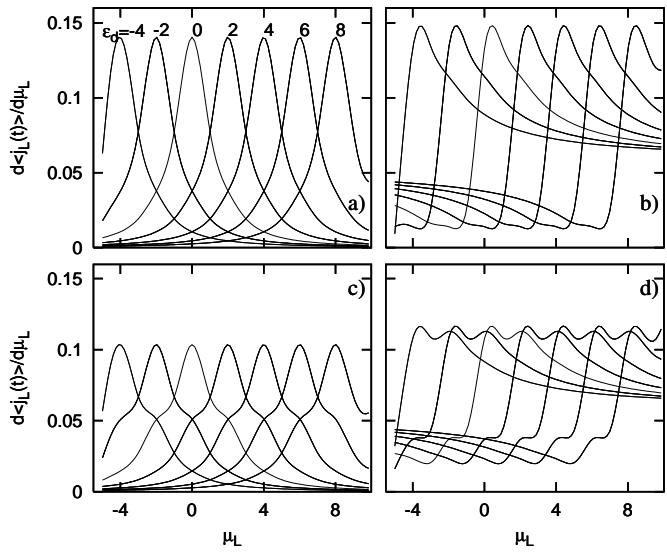}  
\caption {The derivatives of the average current 
with respect to $\mu_L$, $d\langle j_L(t)\rangle/d\mu_L$,
for given values of $\varepsilon_d$ 
(beginning from $\varepsilon_d = -4$ up to $\varepsilon_d = 8$). 
The left (right) panels correspond to $V_{RL} = 0$ ($V_{RL} = 10$) and upper
(lower) panels correspond to $\Delta_L = 2$ ($\Delta_L = 4$).
The other parameters as in Fig. 1.}
\label{7}
\end{figure}

\newpage
\begin{figure} 
\epsfysize=10cm
\includegraphics[angle=0,width=0.9\textwidth]{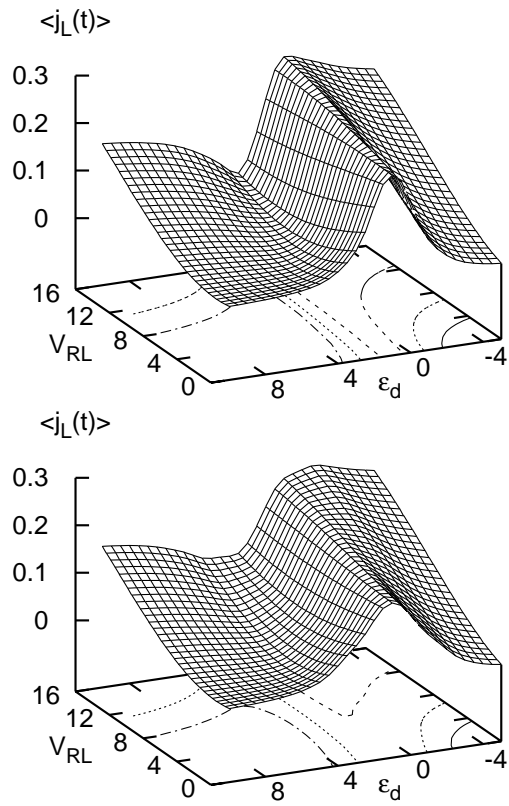}
\caption {The average current $\langle j_L(t)\rangle$ against $V_{RL}$ and $\varepsilon_d$. The upper
(lower) panel corresponds to $\Delta_L = 2$, $\Delta_d = 1$ ($\Delta_L = 4$,
$\Delta_d = 2$). $\mu_L = 2$, $V = 4$.}
\label{8}
\end{figure}

\newpage
\begin{figure} 
\epsfysize=10cm
\includegraphics[angle=0,width=0.9\textwidth]{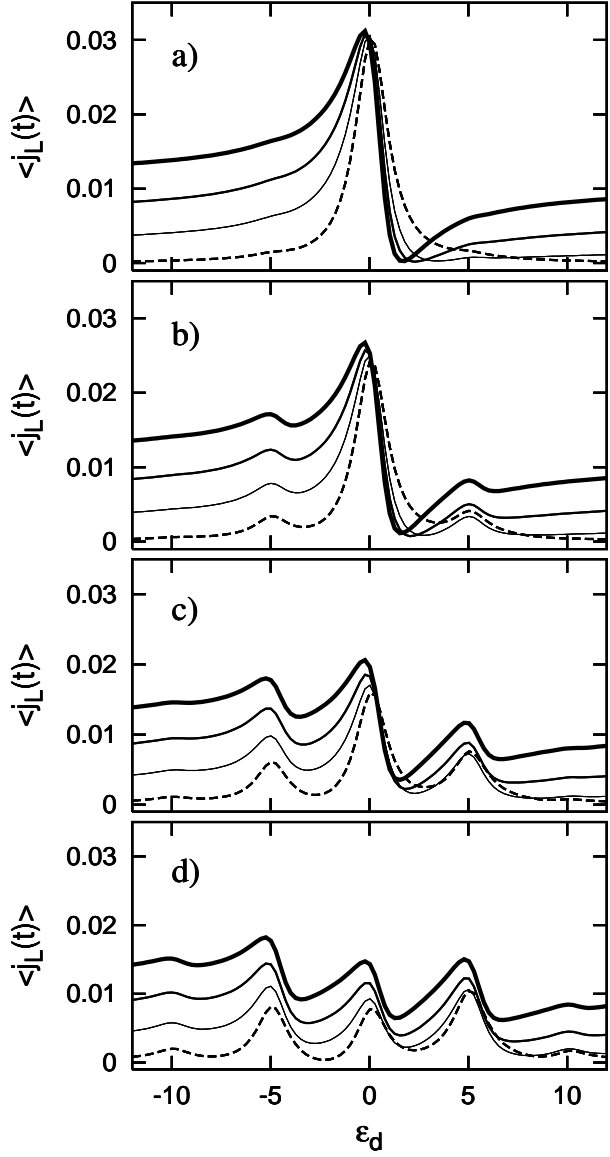}  
\caption {The average current $\langle j_L(t)\rangle$ against $\varepsilon_d$ for
the oscillating QD energy level at $V_{RL}$ = 0, 4, 7, 10 -- broken, thin, thick and
very thick curves, respectively. The panels a, b, c and d correspond to 
$\Delta_d = 1, 3, 5$ and 7, respectively. $\omega = 5$, $\Gamma = 1$, $V = 4$,
$\mu_L = 0.2$, $\Delta_L = \Delta_R = 0$.}
\label{9}
\end{figure}

\newpage
\begin{figure} 
\epsfysize=10cm
\includegraphics[angle=0,width=0.9\textwidth]{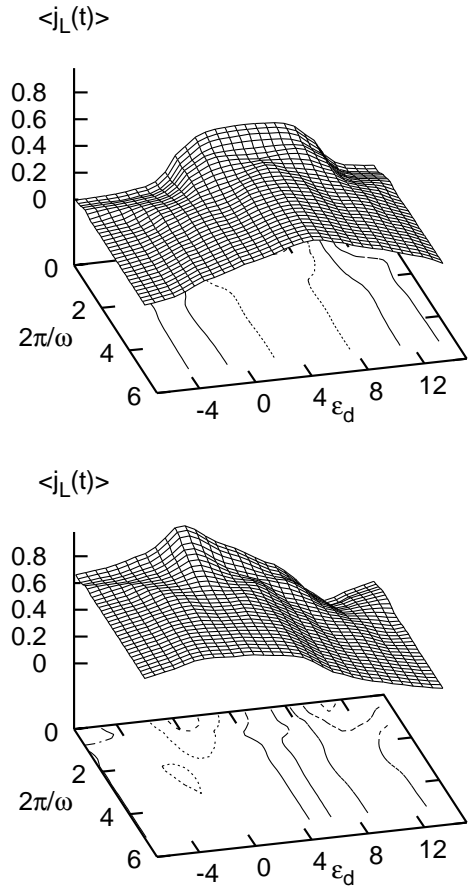}
\caption {The average current $\langle j_L(t)\rangle$ against $\varepsilon_d$ and the
period of the time oscillation of the external field for $V_{RL} = 0$ -- the upper panel
and for $V_{RL} = 10$ -- the lower panel. $V = 4$, $\mu_L = 10$, $\Delta_L = 10$, 
$\Delta_d = 5$, $\Delta_R = 0$.}
\label{10}
\end{figure}

\newpage
\begin{figure} 
\epsfysize=10cm
\includegraphics[angle=270,width=1.0\textwidth]{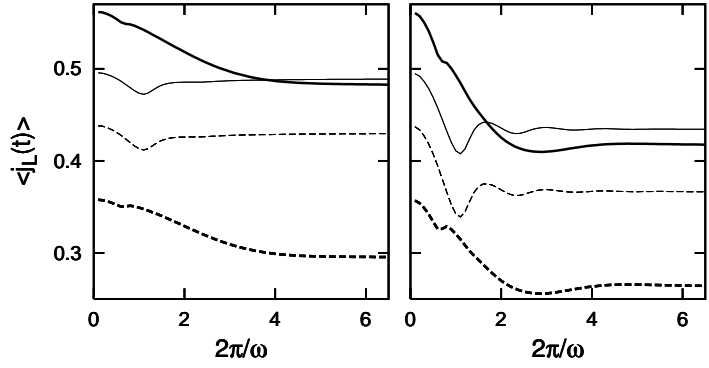}
\caption {The average current $\langle j_L(t)\rangle$ against 
the period of the time oscillation of the external field for $V_{RL} = 0$ -- the
broken curves, and for $V_{RL} = 4$ -- the solid curves. The thin (thick) curves correspond
to $\varepsilon_d = 1$ ($\varepsilon_d = 5$). The left (right) panel corresponds
to $\Delta_L = 5$, $\Delta_d = 2.5$ ($\Delta_L = 10$, $\Delta_d = 5$).
$\Delta_R = 0$, $V = 4$, $\mu_L = 10$.}
\label{11}
\end{figure}

\end{document}